# Crystal Structure and Magnetic Properties of a new two-dimensional S=1 quantum spin system $Ni_5(TeO_3)_4X_2$ (X= Cl, Br)

Mats Johnsson[*], Karl W. Törnroos[**], Peter Lemmens[#] and Patrice Millet[&,+]

*Department of Inorganic Chemistry, Stockholm University, S-106 91 Stockholm, Sweden
**Department of Chemistry, University of Bergen, Allégt. 41, 5007 Bergen, Norway.
[#] Max Planck Institute for Solid State Research, Heisenbergstraße 1, D-70569 Stuttgart, Germany
[&]Centre d'Elaboration de Matériaux et d'Etudes Structurales (CEMES/CNRS), 29 rue Jeanne Marvig, BP 4347, Toulouse cedex 4, France.

**Abstract**

Two new transition-metal tellurium oxy-chlorides of the general formula $Ni_5(TeO_3)_4X_2$ (X=Cl, Br) have been isolated during the investigation of the ternary phase diagram NiO-$NiCl_2$-$TeO_2$. They crystallize in the monoclinic system, space group C2/c, and for the case of $Ni_5(TeO_3)_4Cl_2$ the unit cell is $a$=19.5674(2) Å, $b$=5.2457(1) Å, $c$=16.3084(1) Å, with $\beta$=125.289(1)°. The structure is layered and built up of corner connected $[Ni_5O_{17}X_2]$ entities, made of five nickel(II) octahedra associated by edge and face sharing. The tellurium(IV) atoms are fixed upon the nickel layers. Their lone pairs E and the halogen atoms are packed in a double layer perpendicular to the [010] direction. From the magnetic point of view, this system provides a new 2D S=1 quantum spin system with antiferromagnetic superexchange interaction. The magnetic susceptibility shows anomalies pointing to magnetic ordering phenomena. The observed transition temperatures vary with the interlayer separation.

[+] **Corresponding author** : P. Millet, CEMES-CNRS, Toulouse, France
Tel: 33 5 62 25 78 28; Fax: 33 5 62 25 79 99; e-mail: millet@cemes.fr

**Introduction**

Quantum spin systems are compounds in which long range magnetically ordered phases (ferro- or antiferromagnetic) are suppressed by the effect of quantum fluctuations. These fluctuations arise due to a reduced dimensionality of the system (chains or layers), the small spin (S=1/2 or 1) and/or the effect of magnetic frustration (competition between exchange processes). Very often two-dimensional exchange topologies provide the basis for an effective interplay between competing interactions with remaining spin anisotropies. This leads to rich phase diagrams of short and long range ordered phases in dependence of certain coupling parameters. The search for new materials that could exhibit interesting "non classical" magnetic properties is one of the big challenges in condensed matter physics. Recently compounds containing copper(II) and vanadium(IV) (both elements with S=1/2) have been isolated mainly on the basis of topological considerations aiming to select peculiar spin arrangements (1,2). A careful investigation of e.g. the $V^{4+}$ rich zone of the sodium-vanadium-oxygen phase diagram has, e.g., has allowed us to synthesize and characterize two new low-dimensional phases $\eta Na_{1.286}V_2O_5$ (3) and $Na_2V_3O_7$ (4), the latter compound corresponds to the first quantum spin nanotube system.

Another synthesis strategy that has proved to be successful is the use of so-called lone pair cations like Te(IV), Se(IV),.etc., for which it has already been demonstrated by Galy et al. (5) that the effective volume of the lone pair, denoted E, is approximately the same as the volume of an $O^{2-}$ ion. As a consequence these elements, when mixed with a transition metal in presence of halogen ions , can be regarded as "chemical scissors". A study of the system $CuO-CuX_2-TeO_2$ (X=Cl, Br) resulted in the discovery of two isostructural compounds with the general formula $Cu_2Te_2O_5X_2$ which possess a unique crystal structure formed of isolated tetrahedral clusters of copper(II) (6) and which present 3D magnetic ordering for the chlorine system and evidence of an unconventional magnetic instability for $Cu_2Te_2O_5Br_2$ (7). S=1/2 Kagomé type lattice has also been observed in synthetic analogues of $Cu_3Bi(SeO_3)_2X$ (X=Cl, Br, I) (8).

The aim of this study was to look for new phases showing low dimensional spin couplings in between the magnetic ions in the system $NiO-NiX_2-TeO_2$ (X = Cl, Br, I). To our knowledge only three oxide phases have been described before containing Te(IV) in combination with Ni(II); $NiTe_2O_5$ (9), $Ni_2Te_3O_8$ (10), and $Ni_3(OH)_2(TeO_3)_2$ (11).

This study led to the discovery of two new phases with the general formula $Ni_5(TeO_3)_4X_2$ (X = Cl, Br), that are layered structures with two-dimensional arrangements of $NiO_6$ and $NiO_5X$ octahedra and $TeO_3E$ tetrahedra. In such coordinations the $3d^8$ electron configuration of

nickel(II) usually leads to well-localized S=1 spins and a negligible or quenched orbital momentum. This paper presents the synthesis, structural description and a preliminary analysis of the magnetic properties of these peculiar S=1 quantum spin systems.

**Experimental Section**

The syntheses were made by chemical transport reactions in sealed evacuated quartz glass tubes. As starting materials NiO (Avocado Research Chemicals Ltd.), $NiCl_2$ (>98% Merck, wasserfrei zur synthese), $NiBr_2$ (Stream Chemicals, 99+%), $NiI_2$ (Strem Chemicals, 99.5%) and $TeO_2$ (Strem Chemicals, 99+%) were used. The $Ni_5(TeO_3)_4X_2$ compounds were made from a mixture of NiO : $NiX_2$: $TeO_2$ (X = Cl, Br, I) in a molar ratio 4 : 1 : 4 at 450°C for 90h. The products of the syntheses were characterised in a scanning electron microscope (SEM, JEOL 820) using an energy-dispersive spectrometer (EDS, LINK AN10000).

The syntheses produced orange $Ni_5(TeO_3)_4Cl_2$ and $Ni_5(TeO_3)_4Br_2$ single crystals and powders, respectively. The analogous iodine phase was not obtained as single crystals, however, a brown multiphase powder was obtained indicating the presence of $Ni_5(TeO_3)_4I_2$. For the magnetic susceptibility measurements monophase powders of the $Ni_5(TeO_3)_4X_2$ phases were prepared from the same stoichiometric molar ratios NiO : $NiX_2$ (X = Cl, Br): $TeO_2$ = 4 : 1 : 4, using the same synthesis temperature and reaction times as in the preparation of the single crystal. Before magnetic susceptibility measurements the compounds were characterised by their X-ray powder diffraction patterns (XRD) obtained in a Guinier-Hägg focusing camera with subtraction geometry. Cu-K$\alpha_1$ radiation ($\lambda$ = 1.54060 Å) was used, and finely powdered silicon ($a$ = 5.43088(4) Å) was added as an internal standard. The recorded films were evaluated in an automatic film scanner. Unit cell parameters were refined with the program PIRUM (12).

The single crystals X-ray data were collected on a Bruker-AXS SMART 2K CCD diffractometer equipped with an Oxford Cryostream crystal cooling system. Full reflection spheres were collected by means of 0.3º $\omega$-scans. The data were collected and reduced using SMART and SAINT (13). Gaussian face indexing absorption correction, structure solution, refinement and graphical illustrations were made with SHELXTL (14). Crystal data are reported in Table 1.

Measurements of dc magnetic susceptibility were performed using a Quantum Design SQUID magnetometer on powder samples in fields from 0.01-5 Tesla. The data have been sampled comparing field cooled (FC) as well as zero field cooled (ZFC) measurements. In addition ac magnetic susceptibility has been investigated in a field of $1 \cdot 10^{-2}$ Tesla with driving frequencies $0.1-10^3$ Hz. Detailed results of the latter experiments will be published elsewhere (15).

**Results and Discussion**

*Crystal structure*

The phases $Ni_5(TeO_3)_4Cl_2$ and $Ni_5(TeO_3)_4Br_2$ are isostructural and crystallize in the monoclinic system, space group C2/c. Atomic coordinates, bond lengths, selected interatomic distances and angles are listed in Tables 2, 3 and 4 respectively. No single crystals were obtained for the iodine analogue but indexation of the strongest diffraction lines in the powder X-rays diffractogram gave the following cell parameters for the phase $Ni_5(TeO_3)_4I_2$; *a*=20.766(7)Å, *b*=5.230(1)Å, *c*=16.464(4)Å, *β*=125.79(2)°.

There are three different crystallographic nickel sites. Ni1 (Wyckoff site 4e) and Ni3 (in 8f) are octahedrally coordinated by oxygen atoms while Ni2 (in 8f) is connected to five oxygen and one halogen atoms. The three octahedra are rather distorted, see Tables 3 and 4 and Figure 1. Worth mentioning is the particular connectivity between different octahedra. [Ni1O$_6$] octahedron is connected via face sharing O1-O4-O5 to two [Ni2O$_5$X] octahedra and via edge sharing O4-O5 to two [Ni3O$_6$] octahedra. [Ni2O$_5$X] shares respectively two faces with [Ni1O$_6$] and [Ni3O$_6$] octahedra, the latter via the oxygen atoms O4-O5-O6. This structural entity of general formula [Ni$_5$O$_{17}$X$_2$] is relatively dense with a very short Ni2-Ni3 distance equal to 2.8204(5) Å. The shape of this entity forms a kind of claw which is the building brick from which the Ni layer is made up. As observed in Figure 2, one claw [Ni$_5$O$_{17}$X$_2$] is linked via seven corners to four of its next-nearest neighbors. That forms, due to the two-fold axis and the glide plane c, rows running alternatively along 0y and -0y.

The tellurium atoms Te1 and Te2 present a classical coordination polyhedron, noted [TeO$_3$E] formed by three oxygen atoms and the $5S^2$ lone pair E. However, they exhibit very different bonding schemes. The [Te1O$_3$E] tetrahedron depicted in Figure 3a is quite regular with Te-O distances ranging between 1.878(2) Å and 1.896(2) Å and O-Te-O angles varying between 94° to 99°. They are located above and below the Ni layers and as observed in Figure 3b, each tetrahedron shares three oxygen atoms with four adjacent basic entities [Ni$_5$O$_{17}$X$_2$].

The [Te2O$_3$E] tetrahedron is distorted with a larger range of distances and angles from 1.837(2) Å to 1.914(2) Å and 88° to 98° respectively. As observed in Figure 4 this distortion arises from the linkage of the tetrahedron which actually connects both sides of the same claw [Ni$_5$O$_{17}$X$_2$] via the edge O1-O4 and oxygen atom O2, the lone pair E pointing again alternatively above and below each building block.

A description of the general network is given in Figure 5 by the projection of the structure onto the (010) plane. The lone pairs E and the halogen ions are located toward the space available between the Ni layers and this compound provides a new example in which lone pair-lone pair and lone pair-halogen interactions are observed.

The lone pair positions of Te in these two structures need to be addressed further as the conventional way of calculating (CRYSAT (16)) the coordinates of such lone pairs is to locate them at the apices of regular polyhedra built up on the basis of selected oxygen atoms, *cf.* Table 5. This calculation gives a satisfactory result for the Te1 atoms owing to the consistent Te1-Te1 interlayer distances of 4.81 Å and 5.06 Å for the chlorine and bromide structures, respectively. However, a problem arises for the lone pair of Te2 as the interlayer Te2-Te2 distance is smaller than 3.9 Å in both structures. Since the Te2 atoms point towards one another as seen in Figure 6, the conventionally calculated E2-E2 distance becomes unreasonably short (~ 0.9 Å). One therefore has to consider the possibility that the Te2 lone pair is located closer to the atom centre thereby presenting a more spherical shape. As depicted in Figure 6 there is indeed space in between the Ni layers to accommodate for the E2 lone pairs if the barycenter of the trigonal prism X-X$^{ii}$-X$^{v}$-X$^{vi}$-O4$^{v}$-O1$^{v}$ is chosen as its position, *cf.* Table 5.

*Magnetic properties*

The magnetic susceptibility of Ni$_5$(TeO$_3$)$_4$X$_2$ as a function of temperature shows a Curie-Weiss behavior with the negative Weiss temperatures $\Theta = -50K$ for X=Cl and somewhat smaller values of $\Theta = -46K$ for X=Br and I. This indicates antiferromagnetic correlations between the spin moments of nickel in this family of compounds. Respective experimental data are shown in Figure 7. Assuming a 3d$^8$ – s=1 state the effective moment-related g-factor is determined to be g = 2.21, 2.35 and 1.98, for X=Cl, Br and I, respectively. At least the former two values correspond very well to the above described situation where spin-orbit coupling is still important enough to cause significant deviations of the g values from the free ion values.

At lower temperature kinks in the susceptibility give evidence for collective magnetic ordering. The respective critical temperatures are $T_c$ = 23K, 28K and 30K for X=Cl, Br, and I, respectively, meaning that the transition temperatures increase with increasing distance between the planes. In the inset of Figure 7 the low temperature regime is shown for Ni$_5$(TeO$_3$)$_4$Cl$_2$. A pronounced magnetic field and history dependence is observed comparing field-cooled (FC) and zero-field-cooled (ZFC) data. In larger magnetic fields, B > 3T, the hysteresis is suppressed but a kink at T=11K remains in the susceptibility. For the chloride this effect has an onset slightly below the ordering temperature leading to a large temperature interval ΔT=20K with hysteresis. For the other two compounds ΔT is much smaller, i.e. the onset of the hysterisis is shifted to much lower temperature. We determine the following temperature intervals ΔT = 6K (X=Br) and ΔT < 4K (X=I). These temperature scales are opposite to the dependence of the transition temperatures on stoichiometry. Irreversibility and weak ferromagnetism in antiferromagnets may have different origins, both of micro- as well as macroscopic kinds (17), so further ac susceptibility and specific heat experiments have been performed in the temperature regime around T = 11K where a kink in the susceptibility is observed for the chloride. These experiments do neither give evidence for a frequency dependence or dynamic relaxation nor do they show a large entropy change in the investigated temperature and frequency range (15). Therefore we rule out possible spin glass transitions or a sequence of ordering phenomena related to different Ni sites.

A probable origin of the observed hysteresis effects might be based on the interplay of the random grain orientation of the powder sample with magnetic anisotropies of the spin system. The latter anisotropies can be of single ion origin or due to higher order effects (17). In the present spin system with low symmetry exchange paths and a missing inversion center Dzyaloshinskii-Moriya (DM) interaction exists. This interaction should lead to field-induced moments and individual spin canting with an orientation given by the local exchange topology. The static randomness of these fields is proposed to lead to the observed hysterisis effects. The interplay of DM interaction with low-dimensionality is an interesting topic and has recently been discussed for several quantum spin system. As an example we state here the two-dimensional vanadate K$_3$V$_2$O$_8$ (18). The presented data together with further experiments on Ni$_5$(TeO$_3$)$_4$X$_2$ should allow to draw conclusion on the relevance of interlayer distances for its dimensionality and spin anisotropy.

**Conclusion**

Two new nickel-tellurium-oxy-chlorides having the common formula Ni$_5$(TeO$_3$)$_4$X$_2$ (X = Cl, Br) have been isolated during the investigation of the ternary phase diagram NiO-NiCl$_2$-TeO$_2$. The two compounds crystallize in the monoclinic system, space group C2/c. Ni$_5$(TeO$_3$)$_4$Cl$_2$ have the cell parameters $a$ = 19.5674(2) Å, $b$ = 5.2457(1) Å, $c$ = 16.3084(1) Å, and $\beta$ = 125.289(1)°. Ni$_5$(TeO$_3$)$_4$Br$_2$ have the cell parameters $a$ = 20.2554(11) Å,

$b$ = 5.2498(3) Å, $c$ = 16.3005(9) Å, and $\beta$ = 124.937(1)°. No single crystals were obtained for the iodine analogue but indexing of the strongest powder X-ray diffraction lines gave the following cell parameters for the $Ni_5(TeO_3)_4I_2$ phase; $a$ = 20.766(7)Å, $b$ = 5.230(1)Å, $c$ = 16.464(4)Å, $\beta$ = 125.79(2)°.

The compounds show a layered structure built up of corner connected [$Ni_5O_{17}X_2$] entities, made of five Ni(II) octahedra associated by edge and face sharing. The Te(IV) atoms are fixed upon the nickel layers. Their lone pairs E and the halogen atoms are packed in a double layer perpendicular to the [010] direction.

This system provides a new 2D S = 1 quantum spin system with an antiferromagnetic superexchange interaction. The magnetic susceptibility as a function of temperature shows a Curie-Weiss behavior with negative Weiss constants of roughly $\Theta$ = -50K. Evidence for magnetic ordering phenomena exist at lower temperatures. The transition temperatures as well as the onset temperatures of a divergence of field-cooled and zero-field-cooled susceptibilities vary with the interlayer separation. These effects are attributed to random fields induced by spin canting.

**Acknowledgement**
We would like to thank J. Pommer and A. Tillmans for help with the magnetic characterization and R. Kremer, Ch. Geibel, G. Güntherodt and J. Galy for important discussions.

**Table 1.**
**Crystal Data and Structure Refinement for $Ni_5(TeO_3)_4Cl_2$ and $Ni_5(TeO_3)_4Br_2$.**

| | | |
|---|---|---|
| Empirical formula | $Ni_5Te_4O_{12}Cl_2$ | $Ni_5Te_4O_{12}Br_2$ |
| Formula weight | 1066.85 | 1155.77 |
| Temperature | 153(2) K | 153(2) K |
| Wavelength | 0.71073 Å | 0.71073 Å |
| Crystal system | Monoclinic | Monoclinic |
| Space group | C2/c | C2/c |
| Unit cell dimensions | a = 19.5674(2) Å | a = 20.2554(11) Å |
| | b = 5.2457(1) Å | b = 5.2498(3) Å |
| | c = 16.3084(1) Å | c = 16.3005(9) Å |
| | β= 125.289(1)° | β= 124.937(1)° |
| Volume | 1366.38(3) Å$^3$ | 1420.96(14) Å$^3$ |
| Z | 4 | 4 |
| Density (calculated) | 5.186 Mg/m$^3$ | 5.403 Mg/m$^3$ |
| Absorption coefficient | 15.585 mm$^{-1}$ | 20.234 mm$^{-1}$ |
| F(000) | 1912 | 2056 |
| Crystal size | 0.325 x 0.137 x 0.026 mm$^3$ | 0.20 x 0.19 x 0.03 mm$^3$ |
| Theta range for data collection | 2.55 to 30.51° | 2.45 to 31.51° |
| Index ranges | $-27 \leq h \leq 27, -7 \leq k \leq 7, -23 \leq l \leq 23$ | $-29 \leq h \leq 29, -7 \leq k \leq 7, -23 \leq l \leq 23$ |
| Reflections collected | 11193 | 12157 |
| Independent reflections | 2088 [R(int) = 0.0300] | 2363 [R(int) = 0.0549] |
| Completeness to θ | 99.9 % | 99.7% |
| Absorption correction | Numerical | Numerical |
| Max. and min. transmission | 0.6801 and 0.0963 | 0.9407 and 0.1982 |
| Refinement method | Full-matrix least-squares on F$^2$ | Full-matrix least-squares on F$^2$ |
| Data / restraints / parameters | 2088 / 0 / 106 | 2363 / 0 / 105 |
| Goodness-of-fit on F$^2$ | 1.116 | 0.889 |
| Final R indices [I>2sigma(I)] | R1 = 0.0170, wR2 = 0.0375 | R1 = 0.0239, wR2 = 0.0468 |
| R indices (all data) | R1 = 0.0195, wR2 = 0.0379 | R1 = 0.0358, wR2 = 0.0478 |
| Extinction coefficient | 0.00082(3) | - |
| Largest diff. peak and hole | 1.127 and -0.832 e.Å$^{-3}$ | 1.188 and –1.610 e.Å$^{-3}$ |

**Table 2.**

**Atomic Coordinates (x 10$^4$) and Equivalent Isotropic Displacement Parameters (Å$^2$ x 10$^3$) for Ni$_5$(TeO$_3$)$_4$Cl$_2$ (top) and Ni$_5$(TeO$_3$)$_4$Br$_2$ (bottom). U(eq) is Defined as One Third of the Trace of the Orthogonalized U$^{ij}$ Tensor.**

|     | x       | y       | z       | Ueq   |
|-----|---------|---------|---------|-------|
| Te1 | 6266(1) | 1829(1) | 1377(1) | 7(1)  |
| Te2 | 3521(1) | 7879(1) | 1203(1) | 7(1)  |
| Cl  | 2621(1) | 1915(1) | -937(1) | 14(1) |
| Ni1 | 1/2     | 2575(1) | 1/4     | 11(1) |
| Ni2 | 4089(1) | 2842(1) | 265(1)  | 8(1)  |
| Ni3 | 5095(1) | 7191(1) | 1213(1) | 8(1)  |
| O1  | 4347(1) | 77(4)   | 1304(1) | 9(1)  |
| O2  | 6160(1) | 8377(4) | 2506(1) | 11(1) |
| O3  | 4281(1) | 1204(4) | -702(1) | 9(1)  |
| O4  | 4231(1) | 5099(4) | 1448(1) | 9(1)  |
| O5  | 5410(1) | 3333(4) | 1428(1) | 10(1) |
| O6  | 4035(1) | 6469(4) | -194(1) | 10(1) |

|     | x       | y       | z       | Ueq   |
|-----|---------|---------|---------|-------|
| Te1 | 6216(1) | 1846(1) | 1358(1) | 7(1)  |
| Te2 | 3569(1) | 7833(1) | 1201(1) | 7(1)  |
| Br  | 2628(1) | 1969(1) | -983(1) | 13(1) |
| Ni1 | 1/2     | 2531(1) | 1/4     | 10(1) |
| Ni2 | 4136(1) | 2834(1) | 285(1)  | 8(1)  |
| Ni3 | 5102(1) | 7183(1) | 1224(1) | 7(1)  |
| O1  | 4370(2) | 27(5)   | 1307(2) | 9(1)  |
| O2  | 6122(2) | 8358(5) | 2507(2) | 9(1)  |
| O3  | 4303(2) | 1198(5) | -700(2) | 10(1) |
| O4  | 4252(2) | 5055(5) | 1458(2) | 9(1)  |
| O5  | 5394(2) | 3333(5) | 1432(2) | 9(1)  |
| O6  | 4087(2) | 6447(5) | -177(2) | 10(1) |

**Table 3**

**Bond lengths and interatomic distances (Å) for $Ni_5(TeO_3)_4Cl_2$ and $Ni_5(TeO_3)_4Br_2$.**

|  | $Ni_5(TeO_3)_4Cl_2$ | $Ni_5(TeO_3)_4Br_2$ |
|---|---|---|
| Te1-O3$^{vii}$ | 1.878(2) | 1.875(3) |
| Te1-O6$^{iii}$ | 1.884(2) | 1.879(3) |
| Te1-O5 | 1.896(2) | 1.904(3) |
| Te2-O2$^{i}$ | 1.837(2) | 1.841(3) |
| Te2-O4 | 1.891(2) | 1.886(3) |
| Te2-O1$^{ii}$ | 1.914(2) | 1.915(3) |
| Ni1-O4 (x2) | 1.995(2) | 1.997(3) |
| Ni1-O1 (x2) | 2.067(2) | 2.073(3) |
| Ni1-O5 (x2) | 2.341(2) | 2.331(3) |
| Ni2-O3 | 2.013(2) | 2.0133) |
| Ni2-O6 | 2.025(2) | 2.022(3) |
| Ni2-O1 | 2.059(2) | 2.066(3) |
| Ni2-O4 | 2.140(2) | 2.135(3) |
| Ni2-O5 | 2.164(2) | 2.145(3) |
| Ni2-X | 2.413(1) | 2.564(7) |
| Ni3-O3$^{iii}$ | 2.018(2) | 2.018(3) |
| Ni3-O2 | 2.024(2) | 2.021(3) |
| Ni3-O6 | 2.053(2) | 2.053(3) |
| Ni3-O5 | 2.085(2) | 2.079(3) |
| Ni3-O1$^{ii}$ | 2.170(2) | 2.162(3) |
| Ni3-O4 | 2.227(2) | 2.260(3) |
| Ni1-Ni2(x2) | 2.9958(3) | 2.9811(5) |
| Ni1-Ni3(x2) | 3.2794(4) | 3.2897(6) |
| Ni1-Ni3$^{viii}$ (x2) | 3.5873(4) | 3.5693(4) |
| Ni2-Ni3 | 2.8204(5) | 2.8207(8) |
| Ni2-Ni3$^{viii}$ | 3.3969(4) | 3.3972(5) |
| Ni2-Ni3$^{iii}$ | 3.5789(5) | 3.5763(5) |
| Ni3-Ni3i | 4.4235(4) | 4.4126(5) |

*Symmetry transformations used to generate equivalent atoms:*

*(i) 1-x, y, 1/2-z ;   (ii) x, 1+y, z ;   (iii) 1-x, 1-y, -z ;   (iv) 1-x, 1+y, 1/2-z ;   (v) 1/2-x, 3/2-y, -z ;
(vi) 1/2-x, 1/2-y, 1/2-z ;   (vii) 1-x, -y, -z ;   (viii) x, 1-y, z.*

**Table 4**

**Selected bond angles (°) for Ni$_5$(TeO$_3$)$_4$Cl$_2$ and Ni$_5$(TeO$_3$)$_4$Br$_2$.**

|  | Ni$_5$(TeO$_3$)$_4$Cl$_2$ | Ni$_5$(TeO$_3$)$_4$Br$_2$ |
|---|---|---|
| O3$^{vii}$-Te1-O6$^{iii}$ | 93.78(8) | 94.22(12) |
| O3$^{vii}$-Te1-O5 | 99.19(8) | 99.30(12) |
| O6$^{iii}$-Te1-O5 | 94.96(8) | 94.90(12) |
| O2$^{i}$-Te2-O4 | 98.35(8) | 97.80(12) |
| O2$^{i}$-Te2-O1$^{ii}$ | 95.98(8) | 95.23(12) |
| O4-Te2-O1$^{ii}$ | 88.01(8) | 88.21(12) |
| O4$^{i}$-Ni1-O4 | 96.82(11) | 96.87(17) |
| O4-Ni1-O1 (x2) | 82.03(8) | 82.14(11) |
| O4-Ni1-O1$^{i}$ (x2) | 168.58(7) | 168.05(10) |
| O1-Ni1-O1$^{i}$ | 101.33(11) | 101.27(15) |
| O4-Ni1-O5 (x2) | 70.04(7) | 70.29(10) |
| O1-Ni1-O5 (x2) | 72.23(7) | 72.80(10) |
| O4-Ni1-O5$^{i}$ (x2) | 96.65(7) | 95.61(10) |
| O1-Ni1-O5$^{i}$ (x2) | 121.37(7) | 121.65(10) |
| O5-Ni1-O5$^{i}$ | 160.44(9) | 159.20(15) |
| O3-Ni2-O6 | 96.01(7) | 95.62(11) |
| O3-Ni2-O1 | 106.01(7) | 105.85(11) |
| O6-Ni2-O1 | 154.38(7) | 155.38(11) |
| O3-Ni2-O4 | 163.03(8) | 164.77(12) |
| O6-Ni2-O4 | 76.45(7) | 77.16(10) |
| O1-Ni2-O4 | 78.78(7) | 79.08(10) |
| O3-Ni2-O5 | 93.88(7) | 95.10(11) |
| O6-Ni2-O5 | 89.59(8) | 89.44(11) |
| O1-Ni2-O5 | 76.21(7) | 76.99(11) |
| O4-Ni2-O5 | 71.16(7) | 71.69(11) |
| O3-Ni2-Cl (Br) | 87.17(6) | 87.05(9) |
| O6-Ni2-Cl (Br) | 96.05(6) | 94.84(9) |
| O1-Ni2-Cl (Br) | 97.93(6) | 98.10(8) |
| O4-Ni2-Cl (Br) | 108.54(6) | 106.74(8) |
| O5-Ni2-Cl (Br) | 174.12(5) | 175.02(8) |
| O2-Ni3-O3$^{iii}$ | 78.34(8) | 78.74(11) |
| O3$^{iii}$-Ni3-O6 | 94.41(8) | 93.93(11) |
| O2-Ni3-O6 | 171.31(8) | 171.47(11) |
| O3$^{iii}$-Ni3-O5 | 106.49(7) | 107.14(11) |
| O2-Ni3-O5 | 95.57(8) | 95.82(11) |
| O3$^{iii}$-Ni3-O1$^{ii}$ | 109.28(7) | 109.42(11) |
| O2-Ni3-O1$^{ii}$ | 92.47(8) | 93.23(11) |
| O6-Ni3-O1$^{ii}$ | 85.36(7) | 85.05(11) |
| O5-Ni3-O1$^{ii}$ | 144.21(7) | 143.38(11) |
| O3$^{iii}$-Ni3-O4 | 167.87(7) | 167.27(11) |
| O2-Ni3-O4 | 113.52(7) | 113.77(11) |
| O6-Ni3-O4 | 73.98(7) | 73.77(11) |
| O5-Ni3-O4 | 70.91(7) | 70.43(10) |
| O1$^{ii}$-Ni3-O4 | 73.89(7) | 73.44(10) |
| O6-Ni3-O5 | 91.06(8) | 90.48(11) |

**Table 5.**

**Lone pair positions, Te-Te interlayer distances, Te-E distances and E-E distances for $Ni_5(TeO_3)_4X_2$ (X=Cl, Br)**

|  | $Ni_5(TeO_3)_4Cl_2$ | | | | $Ni_5(TeO_3)_4Br_2$ | | |
| --- | --- | --- | --- | --- | --- | --- | --- |
| E1[a] | 0.7048 | 0.1824 | 0.2181 | E1[a] | 0.6972 | 0.1868 | 0.2157 |
| E2[a] | 0.2783 | 0.7960 | 0.0433 | E2[a] | 0.2868 | 0.7926 | 0.0447 |
| E2[b] | 0.1904 | 0.7471 | -0.0458 | E2[b] | 0.1896 | 0.7486 | -0.0461 |
| Te1 – Te1 | 4.81 Å | | | | 5.06 Å | | |
| Te2 – Te2 | 3.66 Å | | | | 3.84 Å | | |
| Te1 – E1 | 1.32 Å | | | | 1.33 Å | | |
| Te2 – E2 [b] | 1.01 Å | | | | 1.03 Å | | |
| E2[b] – E2[b] | 1.91 Å | | | | 2.01 Å | | |

[a] Lone pair positions calculated from CRYSAT[15], [b] position corresponding to the barycenter of the trigonal prism X-Xii-Xv-Xvi-O4v-O1v, see Figure 6.

**Figure Captions**

**Figure 1**  Ortep representation of the coordination polyhedra of the three independent nickel atoms with atom labels.

**Figure 2**  Projection of the nickel layer along the [100] direction. [NiO$_6$] octahedra are in black, and [Ni$_2$O$_5$X] octahedra are represented as open polyhedra with balls (Ni) and sticks.

**Figure 3**  a) Coordination polyhedron of Te1 atom with its lone pair E1, and b) Connection of [Te1O$_3$E] polyhedron onto the nickel layer.

**Figure 4**  Connection of [Te2O$_3$E] polyhedron onto one building block [Ni$_5$O$_{17}$X$_2$].

**Figure 5**  Projection of Ni$_5$(TeO$_3$)$_4$X$_2$ (X=Cl, Br) onto the (010) plane.

**Figure 6**  Ellipsoid representation of the interlayer space around Te2 atoms. The lone pair E2 is drawn at the position b listed in Table 5.

**Figure 7**  Magnetic susceptibility of Ni$_5$(TeO$_3$)$_4$X$_2$ (X=Cl, Br, I) at B=1 T as function of temperature. The inset shows the field-cooled (FC-full symbols) and zero-field-cooled susceptibility (ZFC-open symbols) of Ni$_5$(TeO$_3$)$_4$Cl$_2$ in the low temperature regime for different fields. The curves overlap in the measurement at B=5 T.

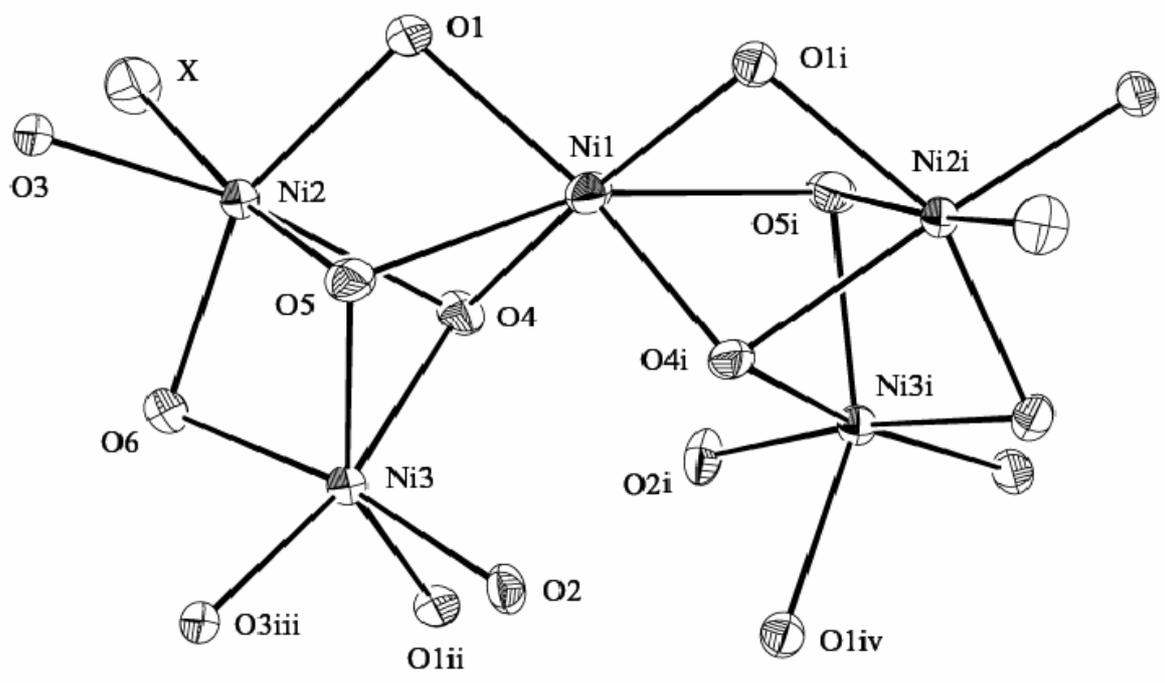

Figure 1    M. Johnsson et al

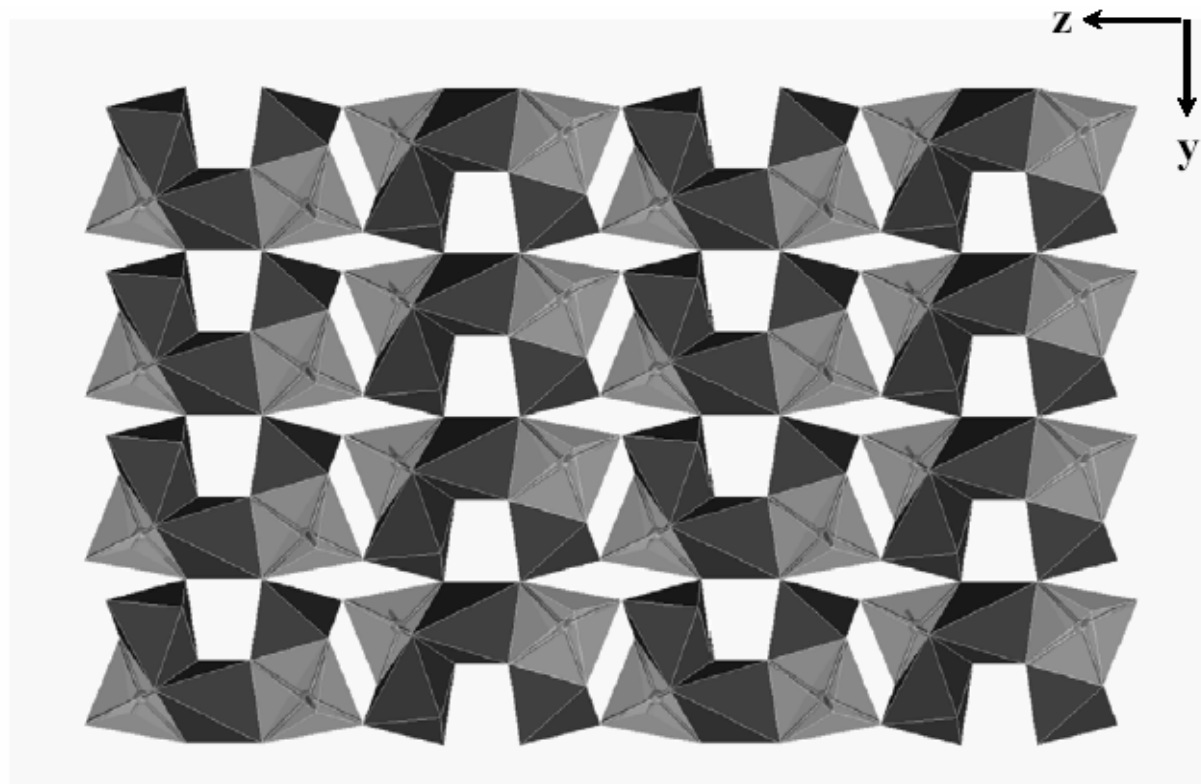

Figure 2    M. Johnsson et al

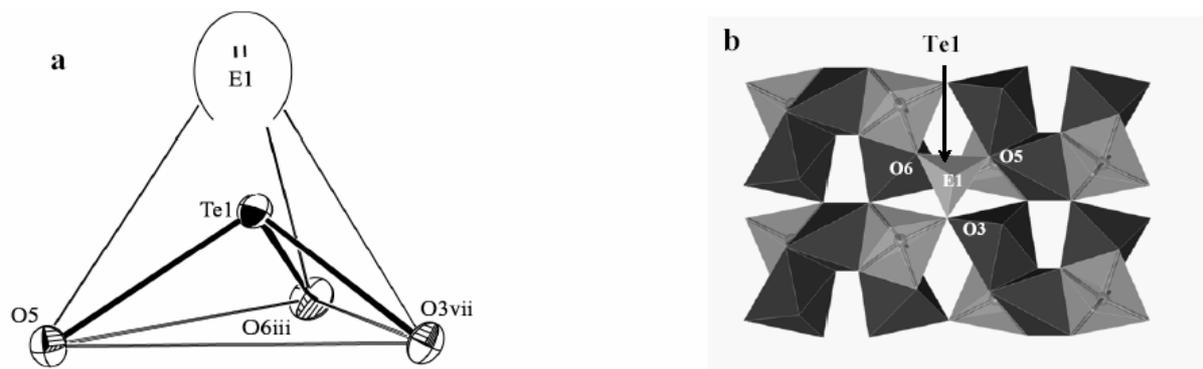

Figure 3    M. Johnsson et al

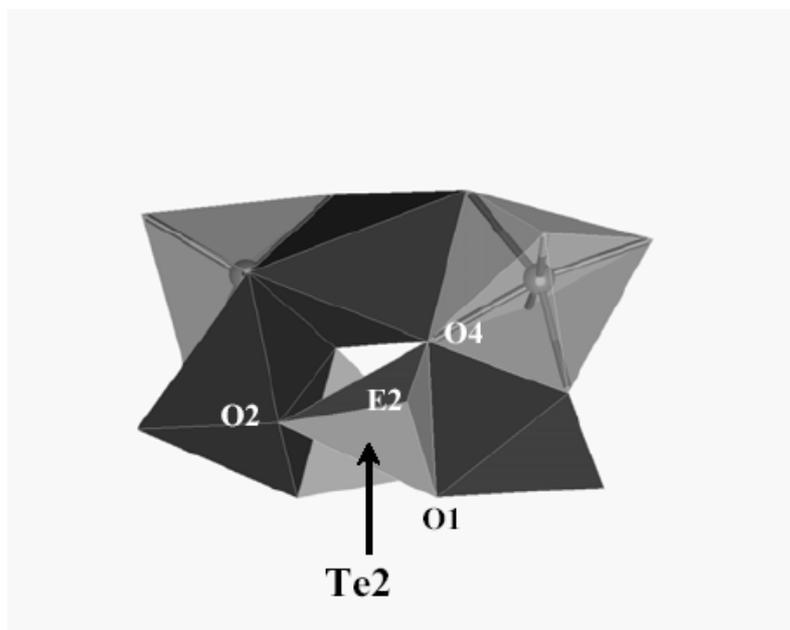

Figure 4    M. Johnsson et al

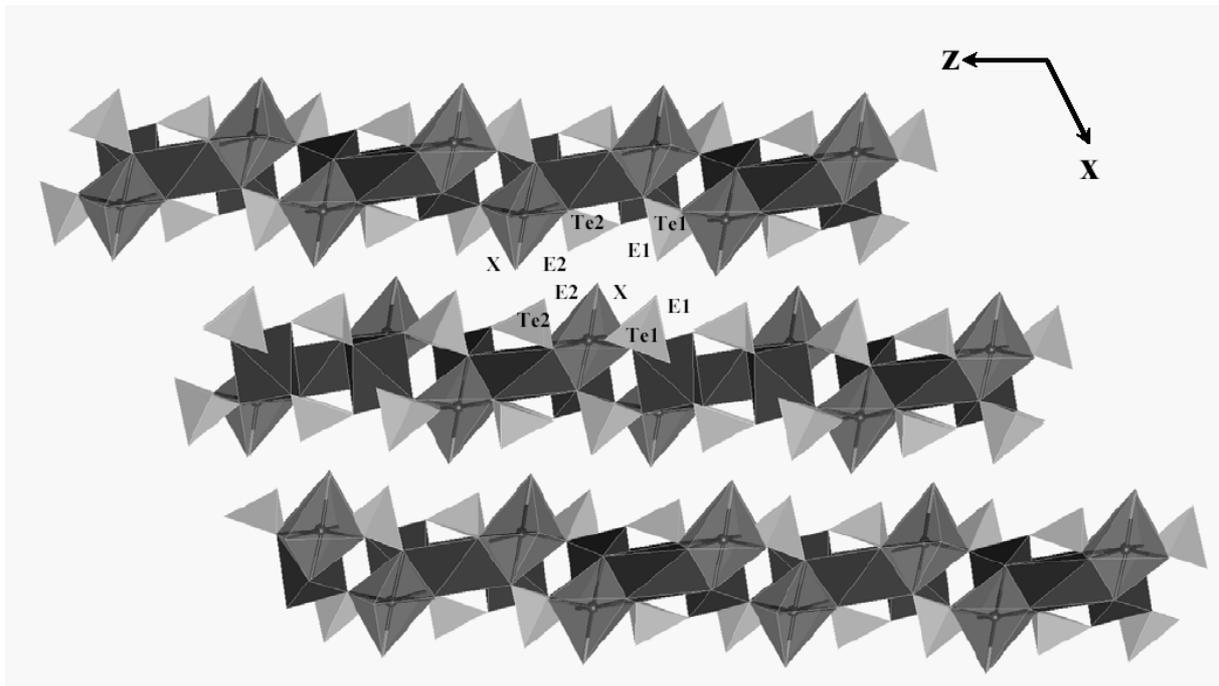

Figure 5    M. Johnsson et al

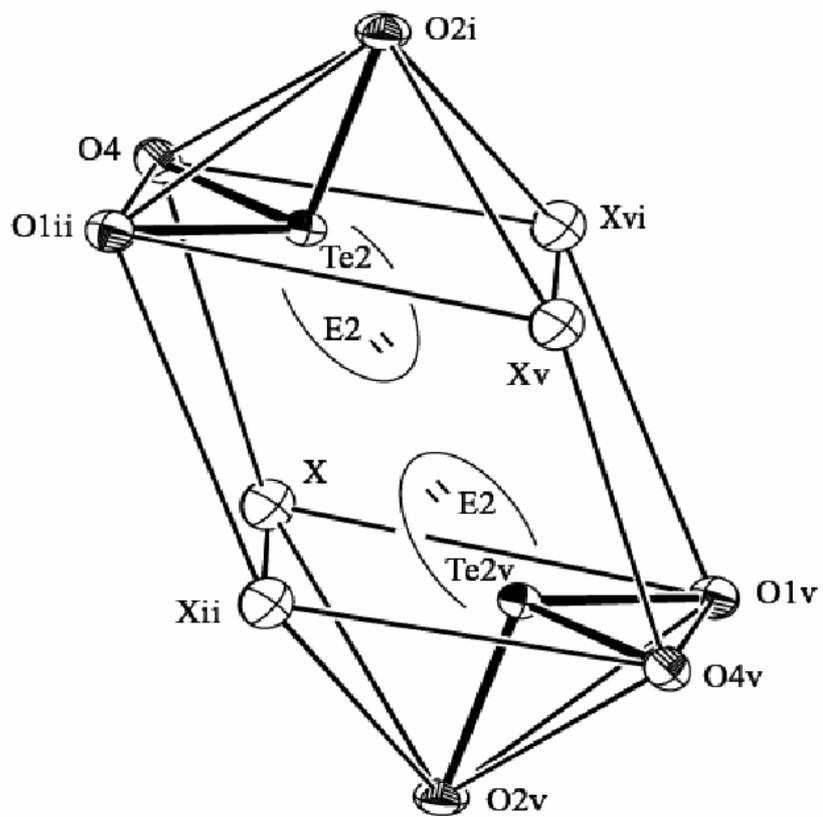

Figure 6    M. Johnsson et al

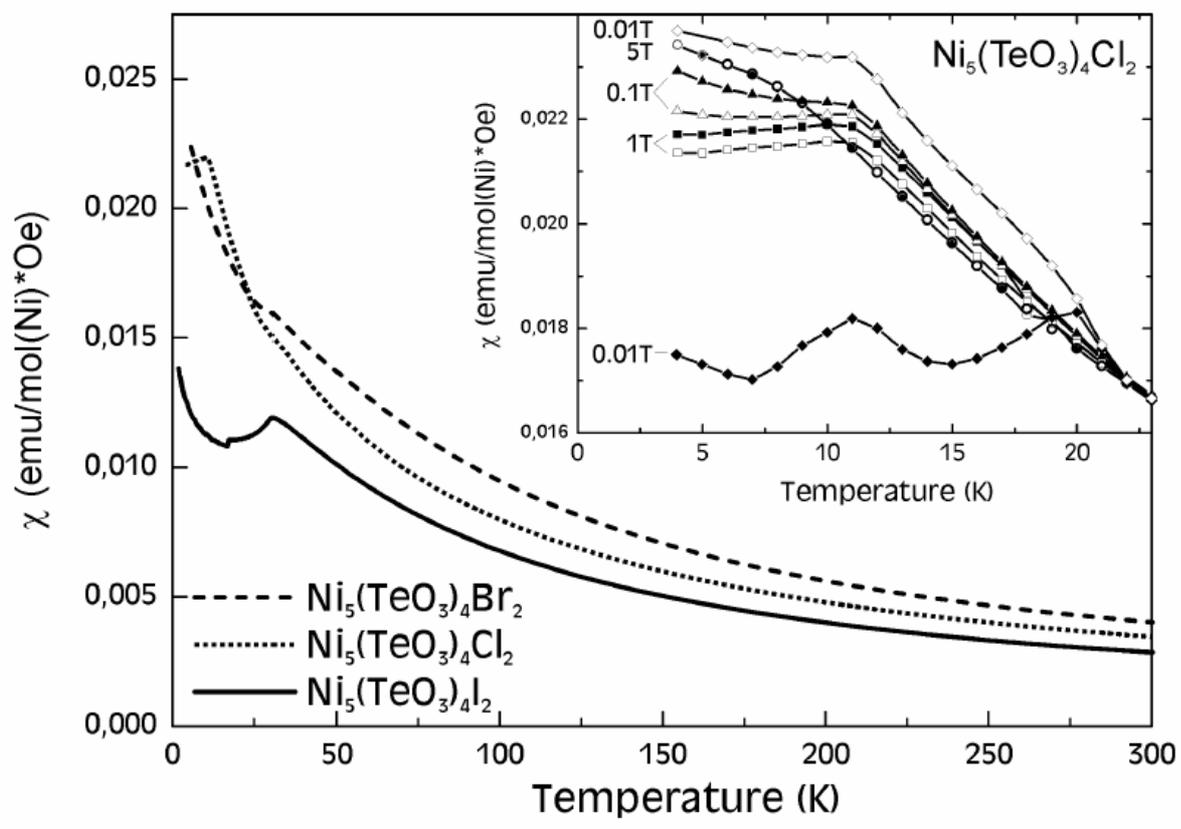

Figure 7    M. Johnsson et al